\documentstyle[11pt,paspconf,epsf]{article}
\begin{document}
%
%
\title{The 1996 Outburst of GRO~J1655--40}
%
%
\author{R.I. Hynes, C.A. Haswell (Also Columbia), 
        R.P. Fender}
\affil{University of Sussex, UK}
\author{C.R. Shrader, W. Chen} 
\affil{NASA Goddard Space Flight Center}
\author{K. Horne, E.T. Harlaftis, K. O'Brien} 
\affil{University of St. Andrews, UK}
\author{C. Hellier}
\affil{University of Keele, UK}
\author{J. Kemp} 
\affil{University of Columbia}
%
%
%
%
\section*{}
During 1996 the soft-X-ray transient (SXT) and superluminal jet source 
GRO J1655--40 was observed in outburst with a multiwavelength campaign 
including {\it HST}, {\it RXTE},  {\it CGRO}\  and ground based facilities.  
The light curves for the outburst are shown in Fig.~1a.  Most striking is 
the sustained X-ray rise, while optical and ultraviolet fluxes were decaying.
This is contrary to standard models of SXT outbursts (e.g.\ Cannizzo, Chen \& 
Livio 1995), and difficult to reconcile with the common view that the optical 
radiation arises from reprocessing of X-rays by the disk and/or secondary.

\begin{figure}
\plottwo{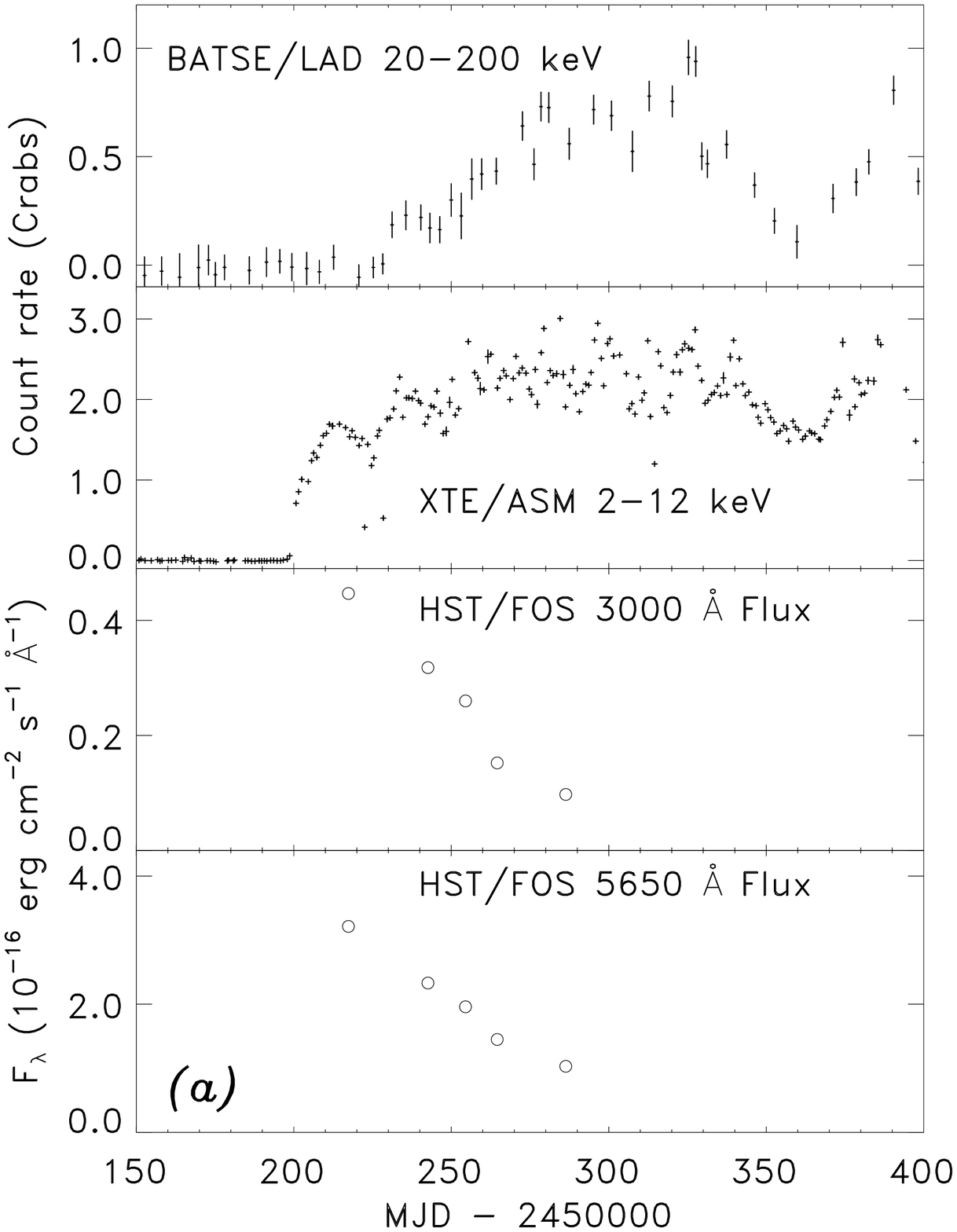}{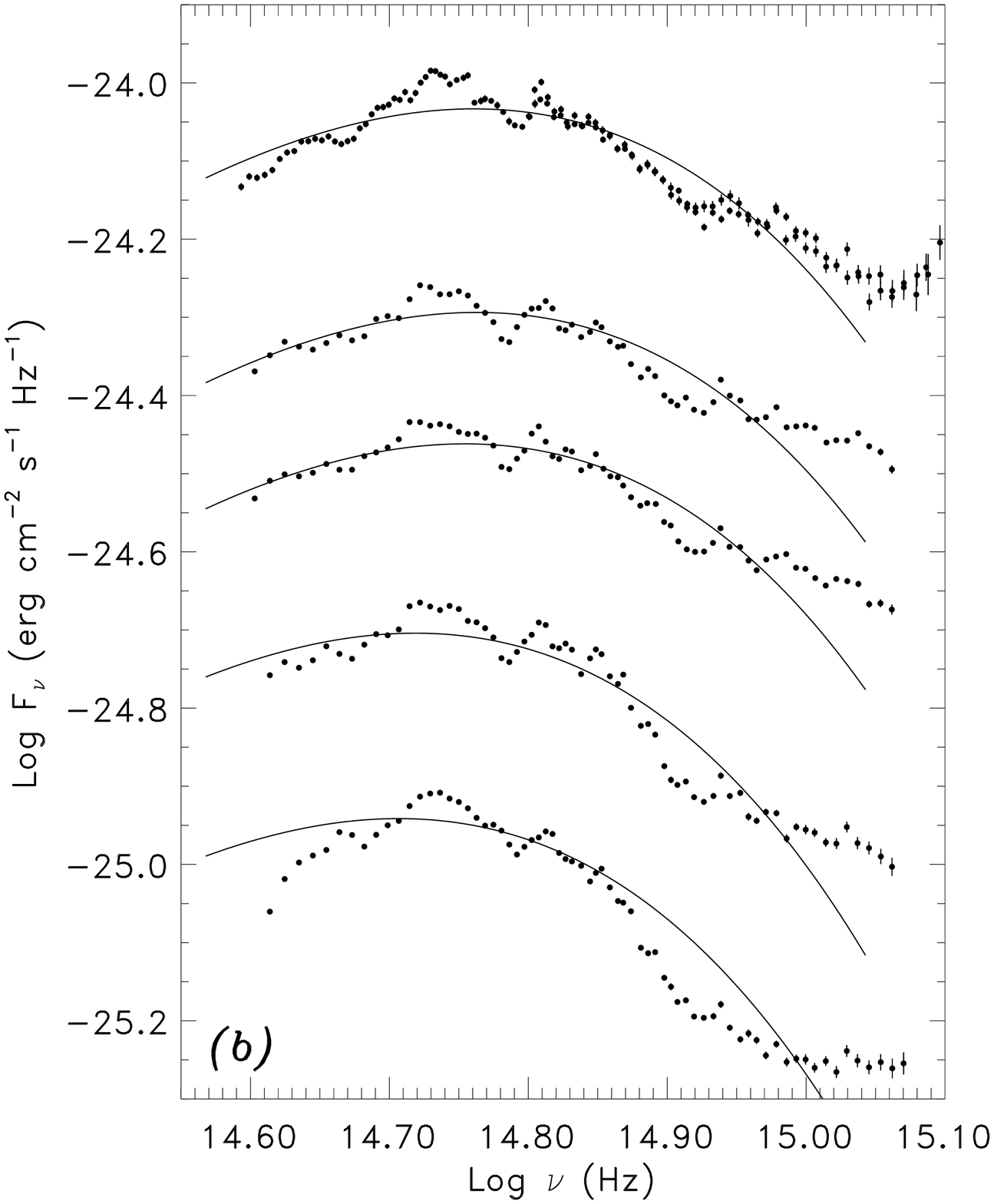}
\caption{a) Light curves of the outburst, spanning 1996 Mar 8 to Nov 13. 
         b) Evolution of the {\it HST} spectra, with best fitting black bodies
            overlaid.  A successive downward shift of 0.1 has been applied to
            each spectrum below the first to separate them clearly.  The 
            wavelength range of this figure is 8500\,\AA\ (left) to 2400\,\AA\
            (right).}
\end{figure}

In the optical, the dereddened {\it HST} spectra (Fig.~1b) do not show the 
$f_{\nu}\propto\nu^{1/3}$ form expected of a steady state accretion disk and 
previously seen in X-ray Nova Muscae 1991 (Cheng et al.\ 1992).  Instead it is 
dominated by a redder component which can be characterized by the 
best-fitting black body.  Roughly we see a fixed temperature emitter 
($T\sim 9000$\,K), shrinking in area from 5.0 to 2.2$\times10^{23}$\,cm$^2$.  
The maximum area is a little larger than the projected disk area, but as 
some of the optical flux will be originating on the secondary, this is 
not inconsistent.

This fixed temperature/shrinking area behavior is suggestive of the inward
propagating cooling wave predicted by the DIM (Cannizzo et al.\ 1995) and the 
temperatures we infer are consistent with material marginally in the hot 
state of that model.  The part of the disk in the hot state is normally 
expected to have a temperature distribution and spectrum similar to a steady
state disk (i.e.\ $f_{\nu}\propto\nu^{1/3}$).  GRO J1655--40, however has a 
longer orbital period (hence larger disk) than most SXTs, so that there exists 
{\em no globally stable steady state solution} for sub-Eddington accretion 
rates.  We would therefore expect a different temperature distribution in 
this case.  The sustained X-ray rise is harder to reconcile with the DIM, but 
may also be a consequence of the large size of the disk.

We have here discussed only one interpretation of the data; an alternative 
source of the optical radiation that we have considered is synchrotron 
emission.  This, and other issues raised by this dataset are discussed in 
Haswell \& Hynes (1997) and a more comprehensive analysis is presented in 
Hynes et al.\ (1997).
%
%
\acknowledgments
RIH would like to thank John Cannizzo, Phil Charles, Andrew King, Kandu
Subramanian and many others for stimulating discussions and Jerome Orosz and 
Roberto Soria for providing their spectra for comparison.
%
%

%
%
\end{document}